\documentclass[amsmath,amssymb,aps,floatfix,showkeys,a4paper]{revtex4}

\usepackage{graphicx}% Include figure files
\usepackage{dcolumn}% Align table columns on decimal point
\usepackage{bm}% bold math
\usepackage{hyperref}% add hypertext capabilities

\usepackage{units}

\begin{document}

\preprint{}

\title{A study on linear and non-linear parton evolution equations}
\thanks{Poster presented by G. C. Penedo in XIV International Workshop on Hadron Physics (18-23 March 2018), Florianop\'{o}lis, Brazil.}%

\author{Penedo, Gilvana C.}
 \email{gilvana.penedo@gmail.com}
\author{Sauter, Werner K.}%
  \email{werner.sauter@ufpel.edu.br}
\affiliation{Grupo de Altas e M\'{e}dias Energias,\\ Instituto de F\'{\i}sica e Matem\'{a}tica,\\ Universidade Federal de Pelotas}%

\date{\today}

\begin{abstract}
In the high energy regime, the proton structure consists of a very large
number of particles called partons (quarks and gluons) that interact with each other, according to the theory of strong interactions, the Quantum Chromodynamics (QCD). Through QCD, the number of partons in the proton is described by the DGLAP equations (linear) and the GLR-MQ equations (nonlinear) of evolution that depend on the kinematic variables $x$ and $Q^2$.  We have studied some analytic and numerical solutions of the GLR-MQ equation. In
order to generate the preliminary results, we used an ansatz for the solution
of the equations of evolution of the gluon distribution, and compared with
results of parametrizations of Parton Distributions Functions (PDFs). In the future, we plan to apply another method to solve the non-linear equations using the Laplace transform.
\end{abstract}

\maketitle

\section{Introduction} 
\begin{figure}
\includegraphics[width=0.4\textwidth]{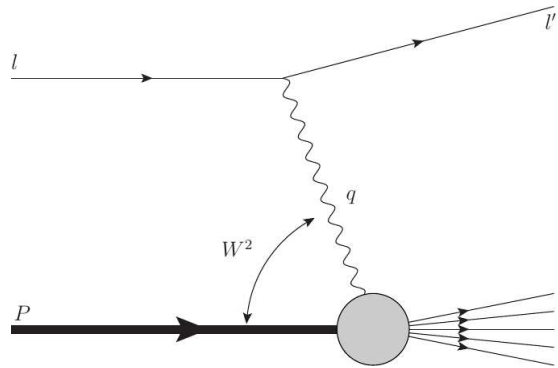}
\caption{Diagram for deep inelastic scattering (DIS) of electron-nucleons. \label{dis}}
\end{figure}

The Standard Model of the Elementary Particles is the theory that describes the constituents of the matter. In this model, we have that the constituents of all matter are fermions (classified into quarks and leptons) and bosons that are intermediary particles, which perform the interaction between fermions. In this work, we will focus in the proton structure, which is formed by quarks and gluons that interact through the nuclear strong force. The theory that studies the strong interaction is the Quantum Chromodynamics (QCD). In QCD, the quarks and gluons have a color charge and are not freely observed in nature. Therefore, the quarks and gluons are only observed as bound states, forming particles called hadrons, which are classified according to the number of valence quarks that constitute them. In particular, the proton is a hadron that have three valence quarks in its structure. The experimental study of the structure of the proton is made through the deep inelastic 
scattering (DIS)(see Fig. (\ref{dis})), which consists of a lepton scattering off a proton, where the lepton emits a photon with virtuality $Q^2$. Other kinematic variables involved in this process are
\begin{equation}
x = \frac{Q^2}{Q^2+W^2-m_N^2},\quad y= \frac{W^2+Q^2-m^2_N}{(l+P)^2-m_N^2},
\end{equation}
where $x$ is the Bjorken variable, related to the fraction of momentum of hadron carried by its
constituents; $y$ is the inelasticity, related to the energy transferred from the lepton to the final state; $W^2$ is the square of center of mass energy of the photon-nucleon system and $m_N$ is the nucleon mass.

\section{Equations of evolution of the parton densities}
In a low energy regime, the proton has its structure established with three valence quarks. At high energies, the proton structure becomes more complex: due to the gluon radiation by the quarks, the gluons are predominant. It is observed that at small $x$ (large $W^2$ with fixed $Q^2$) there is an increase of the density of partons, with a greater increase of the gluon density compared with the quark number. Using QCD, we can describe this behavior with the linear DGLAP evolution equations (Dokshitzer, V. Gribov, Lipatov, Altarelli and Parisi) (for a complete discussion see \cite{devenish2011deep}), which allow us to determine the evolution of the parton distributions if we do not consider the gluon recombination. This coupled system of integral-differential equations is
\begin{equation}
\frac{\partial}{\partial\ln\,Q^2}\left(\begin{array}{c}\Sigma(x,Q^2)\\G(x,Q^2)\end{array}\right) = \frac{\alpha_s}{2\pi}\int^1_x\!dz \left( \begin{array}{cc} P_{qq}(z) & 2n_fP_{qg}(z) \\ P_{gq}(z) & P_{gg}(z) \end{array} \right) \left( \begin{array}{c} \Sigma(x/z,Q^2) \\ G(x/z,Q^2) \end{array} \right),
\end{equation}
where
\begin{eqnarray}
\Sigma(x,Q^2) &=& \sum_{i=1}^{n_f}\left[ xq_i(x,Q^2) + x\bar{q}_i(x,Q^2) \right], \\
G(x,Q^2) &=& xg(x,Q^2),
\end{eqnarray}
and $\Sigma(x,Q^2)$ is the singlet quark distribution, $G(x,Q^2)$ is the gluon distribution and $P_{ij}(z)$ are the splitting functions meaning the probability of a parton $i$ emitting another parton $j$ with a fraction of momentum $z$. 

\section{Saturation and the GLR-MQ Equation}
The DGLAP equation predicts that the gluon distribution function has a strong growth in the small $x$ region. This growth and its theoretical description are related to the violation of the Froissard-Martin limit and the unitarity of the cross section. This is due to the fact that one gluon can emit other gluons. Therefore, another process must be considered in order to reverse the effect described by DGLAP: the recombination of two gluons into one. At a certain energy, the gluons recombine, and this causes the gluon distribution functions to saturate, thus decreasing the density of gluons. The equations of parton evolution modified to include the effect of saturation were obtained by Gribov, Levin and Ryskin \cite{gribov1983semihard} and later by Mueller and Qiu \cite{mueller1986gluon}. To do this, a nonlinear term was included in the parton evolution equations, constituting the so-called GLR-MQ equation. At small $x$, the quark contribution can be 
neglected, thus the GLR-MQ equation for the gluon distribution function $G(x, Q^2) = xg(x, Q^2)$ reads
\begin{equation}
 \frac{\partial G(x, Q^2)}{\partial \ln Q^2} = \frac{\partial G_\mathrm{DGLAP}(x, Q^2)}{\partial \ln Q^2} - \gamma\frac{\alpha_s^2 }{R^2 Q^2}\int_x^1\frac{dy}{y}G^2(y,Q^2), \label{glr}
\end{equation}
where the first term comes from the DGLAP equation (also without the quark contribution) and the second one is the non-linear term. $R$ is an effective radius of gluon interaction and $\gamma = 81/16$.

\section{Solutions for nonlinear evolution equations}
The equations for the gluon distribution $G(x,Q^2)$ can be analytically solved at small $x$ assuming an ansatz for functional dependence of $x$ and $Q^2$, based in the Regge theory: $G(x, Q^2) = x^{-\lambda} H(Q^2)$. After a simple algebra, Eq. (\ref{glr}) became~\cite{devee2014analysis},
\begin{equation}
\frac{dH(t)}{dt}=\gamma_{1}(x)\frac{H(t)}{t}-\gamma_{2}(x)\frac{1}{e^{t}t^2}H^2(t),
\end{equation}
where $t=\log(Q^2/\Lambda^2)$ with $\Lambda^2$ is the QCD energy scale and 
\begin{equation}
\gamma_{1}(x)=\frac{12}{\beta_{0}}\left\{\frac{11}{12}-\frac{N_{f}}{18}+\ln(1-x)+\int_x^1 dz \left[\frac{z^{\lambda+1}-1}{1-z}+\left(z(1-z)+\frac{1-z}{z}\right)z^{\lambda}\right]\right\}.
\end{equation}
and 
\begin{equation}
\gamma_{2}(x)=\gamma\frac{1}{R^{2}\Lambda^{2}}\frac{16\pi^{2}}{\beta_{0}^2} x^{-\lambda}\int_x^1dz z^{2\lambda-1},
\end{equation}
with $\beta_0 = 25/3$. 

The above equation can be solved by usual methods (Bernoulli method) and the solutions are
\begin{equation}
H(t)=\frac{t^{\gamma_1}}{H_{0}-\gamma_{2}\Gamma(\gamma_{1}-1,t)},
\end{equation}
where $\Gamma(a,z)$ is the incomplete gamma function. 
Using the properties for gamma function and after algebraic manipulations, we obtain, accounting for the initial condition,
\begin{equation} \label{eq:sol}
H(t) = \frac{H(t_0)t^{\gamma_{1}(x)}}{t_{0}^{\gamma_{1}(x)}+\gamma_{2}(x)H(t_0)[\Gamma(\gamma_{1}(x)-1,t)-\Gamma(\gamma_{1}(x)-1,t)]}.
\end{equation}
Thus, the distribution of gluons will be given by
\begin{equation} \label{eq:final}
G(x,Q^2) = x^{-\lambda_{g}}H\left(\ln \frac{Q^2}{\Lambda^2}\right).
\end{equation}

\begin{figure}
\includegraphics[width=0.5\textwidth]{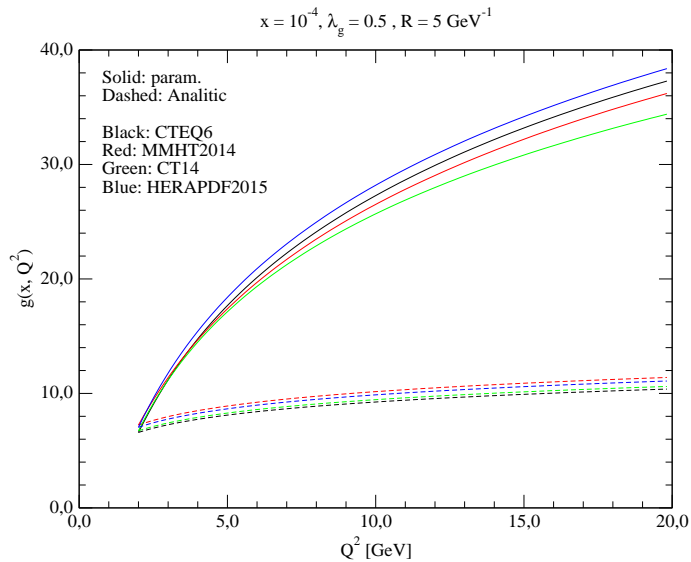} 
\caption{Results for gluon distribution $G(x,Q^2)$ using the analytical result for GLR-MR eq. (dashed curves) compared with LHAPDF interface results (paramaterizations displayed in the legend) as a function of $Q^2$ with $x = 10^{-4}$.  \label{FIG.1} }
\end{figure}

\begin{figure}
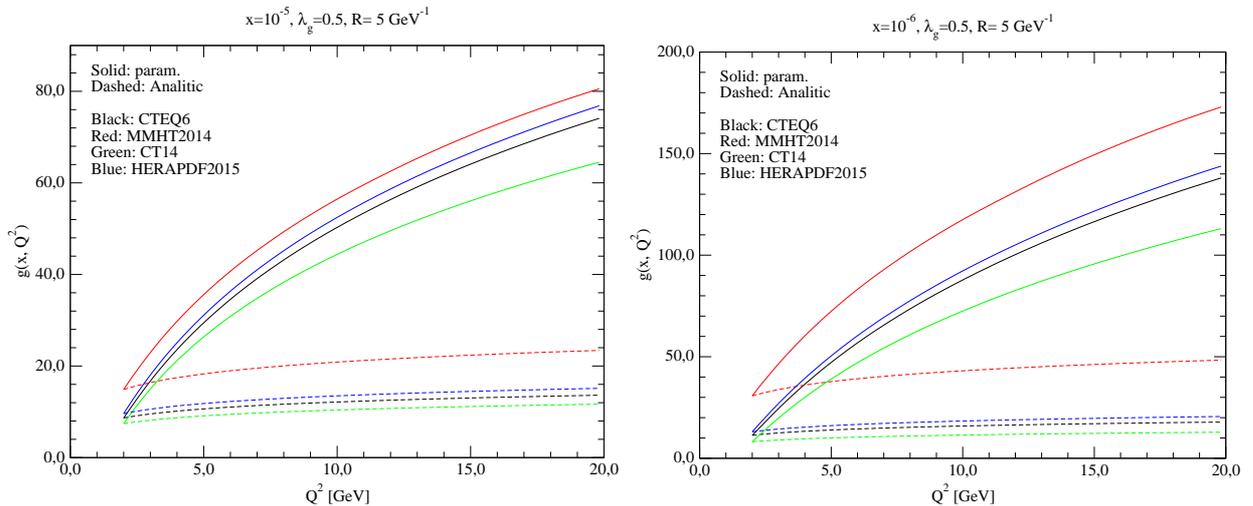

\includegraphics[width=0.45\textwidth]{g555.eps} \includegraphics[width=0.45\textwidth]{g655.eps}
\caption{Same as Fig. (\ref{FIG.1}) but with $x= 10^{-5}$ (left) and $x = 10^{-6}$ (right). \label{FIG.2}}
\end{figure}

From the analytic results, we developed a computational code to calculate the distribution of gluons. 
First, we obtained the initial condition for gluon distribution $G(x,Q^2_0)$ at $Q^2_0 =\unit[1.0]{GeV^2}$ using the values of the CTEQ6, MMHT2014, CT14 and HERAPDF15 parton paramaterizations. Then, we calculated the gluon distribution using the Eq. (\ref{eq:sol}) and (\ref{eq:final}). Using the LHAPDF interface\cite{Buckley:2014ana}, we also calculated the gluon
distributions using the above parton paramaterizations. Both results for the gluon distribution as a function of $Q^2$ are presented in the Figs.~(\ref{FIG.1}) and (\ref{FIG.2}) for different choices of $x$ and fixed values of $\lambda_g = 0.5$ and $R = \unit[5.0]{GeV^{-1}}$.

The results show, as expected, that with the decrease of $x$, the distribution of gluons increases considerably. When $Q^2$ increases, the linear result exhibits rapid growth, while the non-linear result has its increase drastically suppressed. Note that the linear result comes from a parameterization that includes the contribution of quarks and, because of our choice of the initial condition, the linear and non-linear results starts from the same point. We obtain the desirable behavior for the distribution of gluons through a simple functional form that factorized the dependence on $x$ and $Q^2$ for the gluon distribution. However, this form may be too simple, since $\lambda$ may depend on the virtuality of the photon.

\section{Conclusions and future works}

In this paper, we review the results for the analytical solution of equations of the GLR-MQ evolution in a high energy regime through a simple ansatz for the functional form. The final test result is easily implementable and their results show that, in fact, gluon distributions have their growth in the small $x $ substantially suppressed. 

As future works, we intend to study the variation of $\lambda_{g}$ with $Q^2$ and use the Laplace transform method\cite{boroun2013approximate,block2008analytic} to solve the linear and nonlinear evolution equations, comparing their results.

\begin{acknowledgments}
We wish to acknowledge the support of the High and Medium Energies Group of UFPel (GAME/UFPel).
\end{acknowledgments}

\bibliography{apssamp}
\end{document}